\documentclass[manuscript]{aastex61}
\hypersetup{linkcolor=blue,citecolor=blue,filecolor=cyan,urlcolor=magenta}

\newcommand\aastex{AAS\TeX}

\accepted{2017 July 3}
\submitjournal{AJ}

\shorttitle{\aastex\ Multiepoch BVRI Photometry of Luminous Stars in M31 and M33}
\shortauthors{Martin et al.}

\begin{document}

\title{Multi-Epoch BVRI Photometry of Luminous Stars in M31 and M33}

\correspondingauthor{John C. Martin}
\email{jmart5@uis.edu}

\author[0000-0002-0245-508X]{John C. Martin}
\affil{University of Illinois Springfield \\
One University Plaza, MS HSB 314 \\
Springfield, IL, 62703, USA}
\author[0000-0003-1720-9807]{Roberta M. Humphreys}
\affil{Minnesota Institute for Astrophysics \\
University of Minnesota\\
116 Church St. SE\\
Minneapolis, MN, 55455, USA}

\collaboration{(Minnesota Luminous Stars In Nearby Galaxies)}



\begin{abstract}

We present the first four years of BVRI photometry from an on-going survey to annually monitor the photometric behavior of evolved luminous stars in M31 and M33.  Photometry was measured for 199 stars at multiple epochs, including 9 classic Luminous Blue Variables (LBVs), 22 LBV candidates, 10 post-RGB A/F type hypergiants, and 18 B[e] supergiants.  At all epochs the brightness is measured in V and at least one other band to a precision of 0.04 -- 0.10 magnitudes down to a limiting magnitude of 19.0 -- 19.5.  Thirty three (33) stars in our survey exhibit significant variability, including at least two classic LBVs caught in S Doradus type outbursts.  A hyper-linked version of the photometry catalog is at \href{http://go.uis.edu/m31m33photcat}{http://go.uis.edu/m31m33photcat}.

\end{abstract}

\keywords{galaxies:individual(M31,M33) -- stars:massive -- supergiants}



\section{Introduction} \label{sec:intro}

The upper HR Diagram is populated by a number of evolved massive  stars of different types.  They demonstrate mass loss and instabilities in varying degrees ranging from normal mass losing OB supergiants to high mass loss irregular events associated with Luminous Blue Variables (LBVs).  A recently recognized class of luminous transients, sometimes called supernova impostors \citep{2012ASSL..384..249V}, are associated with non-terminal giant massive stars eruptions (i.e the giant eruption of $\eta$ Car) and possibly with the LBVs or S Doradus variables.

Very little is known about the progenitors of the SN impostors or the evolutionary state of many classes of high luminosity stars.  Expanding the sample of candidates together with building a spectroscopic and photometric  baseline of their behavior and physical parameters will yield more insight into these giant eruptions. 

Surveys in the Milky Way are limited by interstellar extinction and uncertain distances. For this reason we initiated a spectroscopic survey to better categorize and identify the luminous star populations in several nearby galaxies.  In a series of papers we have presented the results for M31 and M33 \citep{paper1,paper2,paper3,paper4}.  The target list was derived from recent surveys for luminous and emission line stars from \citet{LGGS}, \citet{2010AstBu..65..140V} and an unpublished $H\alpha$ survey of M33 by Kerstin Weis.  Our survey has cataloged all the known classical LBV stars and numerous LBV candidates and emission-line objects in both galaxies \citep{paper2,paper4}

Since September 2012 we have imaged fields containing the spectroscopic survey targets.  The photometry in this work builds on the \citet{LGGS,LGGS2} survey of M31 and M33 (see Section \ref{sec:phot}).  Unlike their work, our survey is not intended to catalog every star down to a magnitude limit in both galaxies at a single epoch.   It is instead a targeted survey of the most luminous stars and emission line stars to monitor their variability over the past five years and into the foreseeable future.

There are numerous photometric surveys of M31 and M33 that have sought to discover and characterize stellar variability.  Most have focused on the discovery of Cepheid and RR Lyrae variables and consequently have had a shorter duration and quicker cadence than this survey.  DIRECT \citep{direct1,direct2} and WFCAM/INT \citep{2004ASPC..318..261R} each monitored M31 for three years identifying and observing detached eclipsing binaries and Cephieds.  \citet{2006MNRAS.371.1405H} built a comprehensive catalog of variables in M33 with 27 nights over just 17 months.  POINT-AGAPE \citep{pointagape}, WeCapp \citep{wecapp}, POMME \citep{pomme} and Pan-Starrs/Pandromeda \citep{pandromeda} each lasted 1--3 years and focused on micro-lensing events along the line of sight to M31.  \citet{pandromeda} demonstrated that surveys with shorter duration and cadence, may be useful for identifying candidate LBVs but none of these preceding surveys were intended to build a long-term baseline into the future.  At four years duration and counting, our survey has already surpassed the longest contiguous duration of previous efforts.  

Section \ref{sec:phot} describes how the survey is conducted and the photometry is measured from the images, including analysis of the errors and checks on the reliability of the catalog.  Section \ref{sec:discussion} includes a brief discussion of significant results.  The data are presented in Table \ref{tab:goodtarg} (with notes in Table \ref{tab:targnotes}), available in full online.  We also host a hyper-linked version of the catalog at \href{http://go.uis.edu/m31m33photcat}{http://go.uis.edu/m31m33photcat}.  

\section{Photometry} \label{sec:phot}
\subsection{The Images}
Images were obtained of M31 and M33 using an Apogee U42 CCD Camera with a back-illuminated E2v CCD42-40 chip on the F/13 20-inch (0.51 m) telescope at the University of Illinois Springfield Henry R. Barber Research Observatory near Pleasant Plains, IL.  Each image covers a square area about 19.4 arc-minutes on a side with a pixel scale of 0.57 arcseconds per pixel.  The telescope was pointed to efficiently cover the desired targets in both galaxies and ensure useful overlap between images in different epochs.  Imaging typically took place from late-July to January when M31 and M33 are at least 40 degrees above the horizon for many hours.

Images were exposed in four high-throughput broad-band filters manufactured by Astrodon:  Johnson B and V, and Cousins R and I.  The B filter does {\em not} have the red-leak present in most Astrodon B filers manufactured prior to 2013.  Images were taken in V at every epoch except one in 2012.  In 2012 many fields were also imaged in R.  Imaging in the B filter started in 2013 and imaging in the I filter started in 2015.  Table \ref{tab:obs} gives a record of the 199 images of M31 and 77 images of M33 included in this work.

All images have an exposure time of 600 seconds.  They are bias-corrected, dark-subtracted and flat-fielded.  Bias corrections were made from a bias over-scan in each individual image.  A master-dark was composed each night from at least three independent dark frames.  From 2012 through 2015 flat-fielding was done with dome flats taken immediately following the images.  Beginning in 2016, flat-fielding was done with sky flats taken during twilight on the same night.  There is some low-level fringing in the I-band images (due to night sky OH emission) which has not been removed.  WCS solutions were computed for each image using the astronmetry.net plate solving engine \citep{2010AJ....139.1782L} and hand-checked for accuracy.

Almost all of the images were obtained under dark, steady, transparent atmospheric conditions with no Moon.  Three images in May 2017 were obtained in twilight to confirm an outburst of AF And. The average width and shape of the PSF are most influenced by seeing and telescope tracking.  There is negligible variation of a focused PSF across the field.  The seeing measured from the FWHM of the PSF typically ranged from 3--5 arcseconds (Table \ref{tab:obs}).  

\subsection{PSF Fitting with DAOphot}

In the images, some of the targets are crowded by their neighbors to the point where their point-spread-functions (PSFs) overlap significantly, making it difficult to obtain satisfactory brightness measurements with aperture photometry.  In those crowded fields, PSF fitting photometry performs better.  We experimented with several software packages and chose the DOAphot PSF fitting package in IRAF \citep{1987PASP...99..191S}.  Relative to the other PSF fitting software we tested, DAOphot was able to model the PSFs and the background of our images more reliably and with less fine tuning.  Tests with ensemble photometry on uncrowded fields produced comparable results using either DAOphot PSF fitting or traditional aperture photometry.  

The star PSF for each image was modeled by DAOphot using a elliptical Gaussian.  Parameters for the PSF (and their first order derivatives with respect to position on the image) were determined using no fewer than 20 bright stars hand-selected from those DAOphot suggested as the best in each image. 

The stars in each image were identified using a combination of the daofind tool and LGGS catalog \citep{LGGS2}.  The LGGS has carefully identified all resolved and blended sources to a level fainter than the depth of our images.  But the LGGS also does not include some of the foreground stars which are useful as photometric comparisons.  So as a first pass DAOfind selected sources matching the gross PSF characteristics for the image at least $4\sigma$ above the modeled background.  The stars found by DAOfind were compared with the LGGS catalog down to a limiting magnitude estimated for that image (with consideration made for the combined brightness of stars blended with neighbors less than 1.0 arcseconds apart in the LGGS).    

Stars in common to both the DAOfind list and LGGS were retained as acceptable for PSF fitting.  Most stars above the faint limit of the image that are present in the LGGS and not identified by DAOfind were added to the list.  In very densely crowded fields (i.e. NGC 604 and the nucleus of M33) some LGGS stars above the brightness limit were excluded to alleviate intense confusion.  Stars found in the first pass by DAOfind but not in the LGGS were added if they were brighter than 15th magnitude.  Stars fainter than 15th magnitude from the first pass of DAOfind that are not in LGGS were considered carefully for inclusion on a case by case basis.  Stars brighter than 12th magnitude are overexposed in the images.  Those and others with saturated pixels in their PSF (i.e. very bright) were flagged and ignored.  

\subsection{Ensemble Photometry}
Instrumental magnitudes (I) computed from the PSF fit were transformed to the standard magnitude system (M) via the equation:

\begin{equation}
M = I + Z + K*({\delta}C)
\end{equation}

The zero-point (Z) was computed for each image as the unweighted average of the difference between standard and instrumental magnitudes for an ensemble of comparison stars. ${\delta}C$ is the color of the target subtracted from the average color of the comparisons in the ensemble used to compute Z.  The color term (K) for the telescope/filter/camera combination was determined from separate observations of star cluster M67 done at least once a year.  We employed a time average of the K terms determined over the span 2012--2016 to increase their precision because there has been no significant change in those terms over that time (Table \ref{tab:trans}).

\startlongtable
\begin{deluxetable*}{CRCC}
\tablecaption{Color Dependent Photometric Terms for Barber 20-inch Telescope\label{tab:trans}}
\tablecolumns{4}
\tablenum{2}
\tablewidth{0pt}
\tablehead{
\colhead{Coefficient}& \colhead{Value}&
\colhead{STD Error} & \colhead{$R^2$ of Fit} 
}
\startdata
$K_B{(B-V)}$&+0.038&0.013&0.26\\
$K_V{(B-V)}$&-0.016&0.009&0.16\\
$K_V{(V-R)}$&-0.029&0.016&0.12\\
$K_R{(V-R)}$&-0.032&0.015&0.09\\
$K_V{(V-I)}$&-0.018&0.010&0.16\\
$K_I{(V-I)}$&-0.028&0.010&0.12\\
\enddata
\end{deluxetable*}

We used the method of \cite{2005JAVSO..34...76T} to select comparison stars from the AAVSO Photometric All Sky Survey DR9 (APASS)\citep{2014CoSka..43..518H,APASS}.  APASS is an all-sky photometric catalog covering stars between 7 -- 17 magnitude.  The photometric comparisons selected from APASS had to meet the following criteria:
1) fainter than 12 mag, 2) have colors $0 < (B-V) < 0.8$, 3) have published magnitudes in B, V and at least one other band 4) were separated from any neighboring stars by at least 5 arcseconds and 5) small magnitude errors relative to others stars in the same field.  Ideally the comparison stars selected for each field would span the same range of brightness as the targets.  However, most of the comparison stars selected were between 13 -- 16 magnitude because it was difficult to identify suitable comparison stars fainter than 16 mag in APASS fitting our criteria.    

Only comparison stars with DAOphot PSF fits free of significant irregularities (i.e. due to cosmic ray hits or other artifacts) were included when calculating the photometric zero point for an image.  The photometric solutions for each image were also checked for comparison stars with significant systematic errors in their published brightness. Stars were rejected as comparisons on any image if they were consistently (across most images and epochs) more than 2.5 standard deviations offset from the calculated photometric zero point.  

Among the 199 images of M31 there are 6 to 35 (15 on average) comparison stars identified in each field.  On the 77 M33 images, there are between 5 and 10 (7 on average) identified on each.  The data for the comparison stars are listed in Table \ref{tab:comps}.

Occasionally, there is some overlap between images taken on the same night.  When a target was measured on more than one image in the same filter on the same night the independent measures have been averaged into a single reported magnitude (flagged with an "A" in Table \ref{tab:goodtarg}).  The error given for averaged magnitudes is the sum of the individual errors added in quadrature and divided by the number of independent measurements.  Steps were taken to resolve discrepancies in the few cases where there is a large (1.5 x the error) difference between two or more measurements of a target on the same night.  The number of these instances was reduced to almost zero by rejecting measurements of any target within 3 PSF FWHM of the detector edge.

\subsection{Magnitude Errors} \label{sec:errors}
The measured magnitudes were influenced by a combination of the errors in the instrumental magnitude (computed by DAOphot), the error in the zero-point and blending with other nearby stars.  Blending with neighbors, which did not affect all targets, is discussed separately in Section \ref{sec:blend}.  There is also some contribution to the error from uncertainty in the color term (K).  However, those errors are very small since those terms are stable and have been determined to high precision.

\begin{figure}
\figurenum{1}
\label{zeroerror}
\includegraphics[scale=0.33]{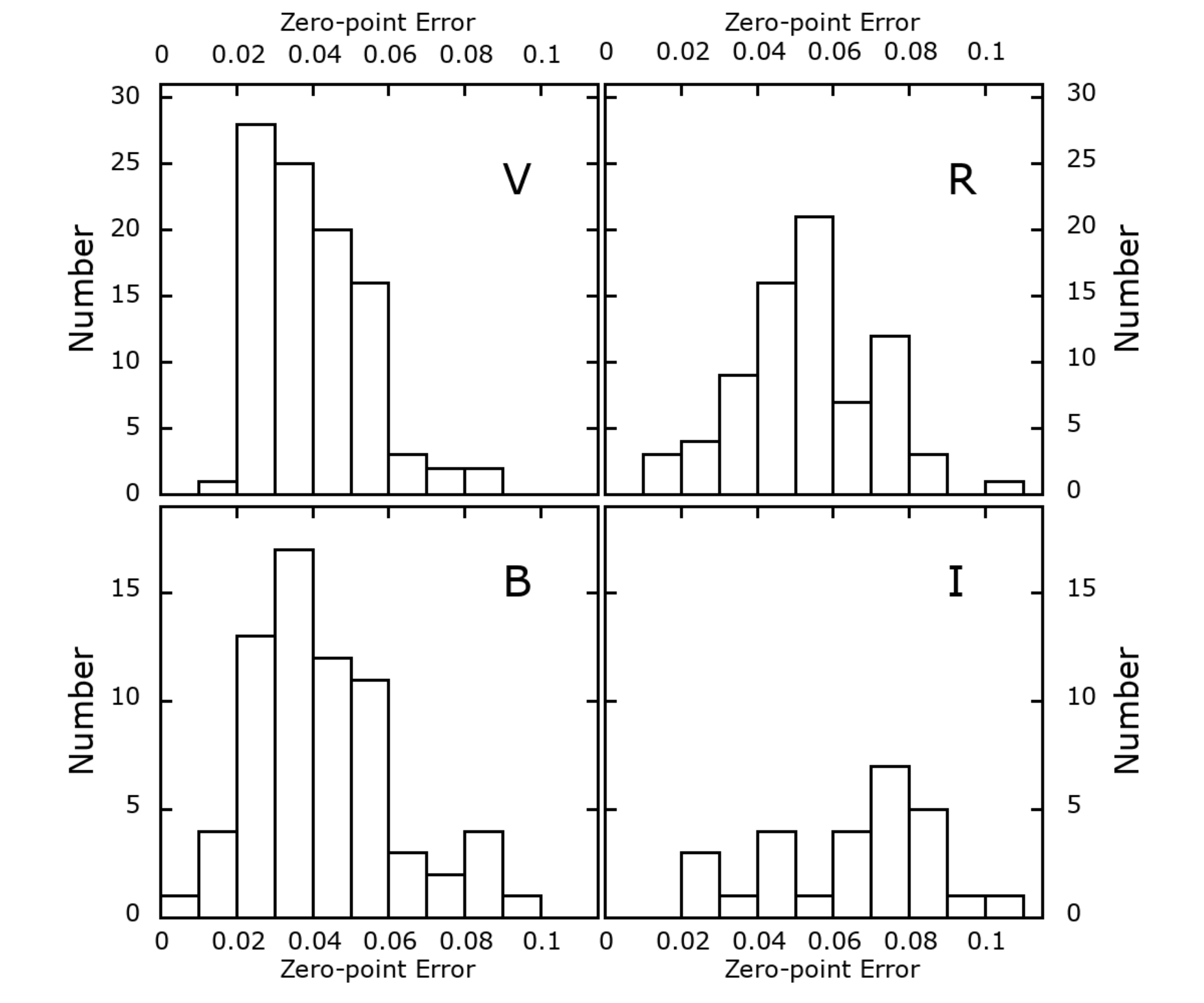}
\caption{Histograms of the photometric zero point errors in the images grouped by filter.}
\end{figure}

The error in the zero-point was estimated from the standard deviation of the mean of the ensemble of comparisons for that image.  Typical zero-point errors are 0.02 to 0.06 magnitudes in B and V (Figure \ref{zeroerror}).  The errors in the R and I tend to be larger because there is typically more uncertainty in the published comparison stars magnitudes in those filters.

\begin{figure}
\figurenum{2}
\label{allmagerr}
\includegraphics[scale=0.33]{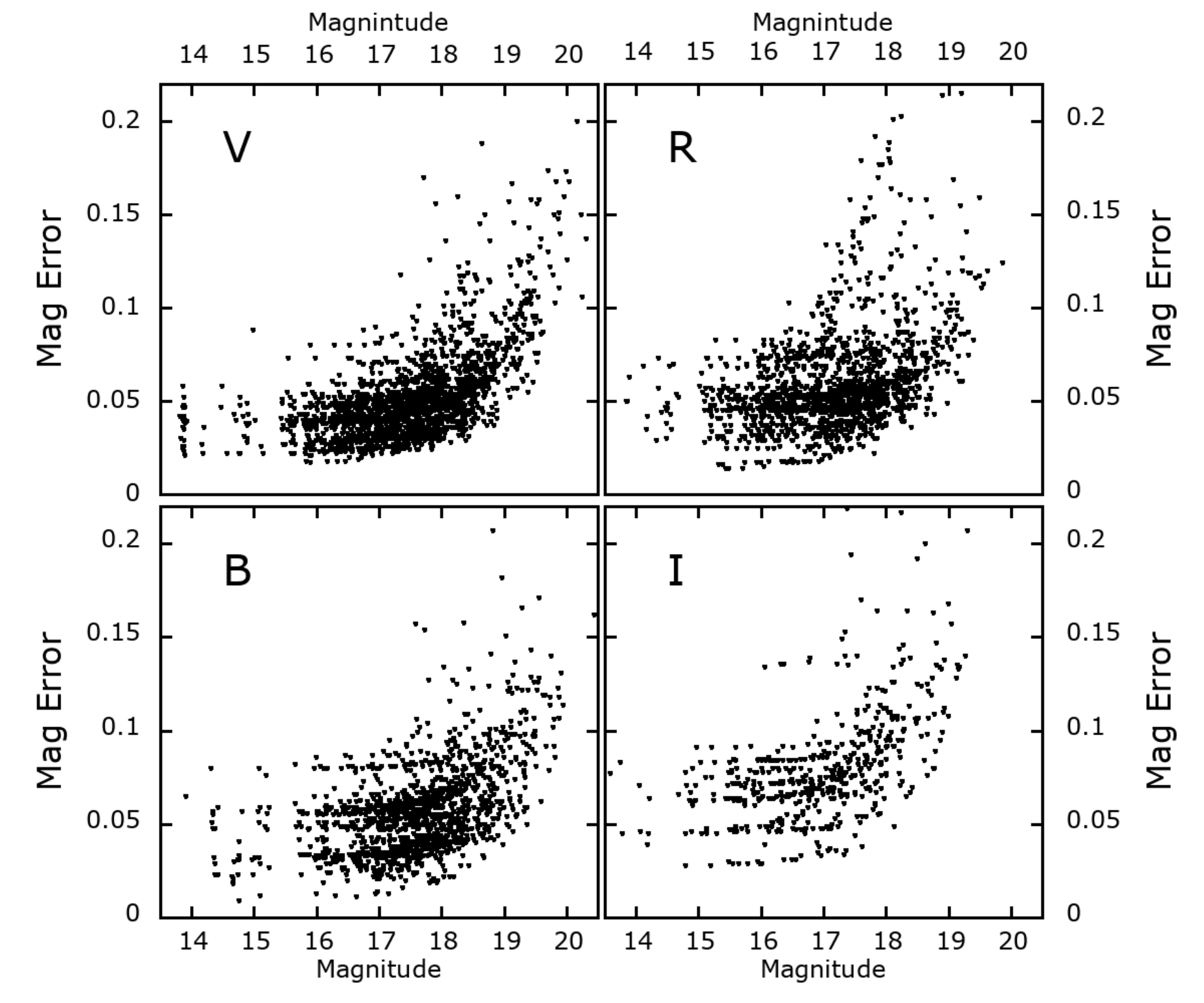}
\caption{Plots of the total error in the magnitude (including zero-point error) grouped by filter.}
\end{figure}

The total magnitude error is plotted in Figure \ref{allmagerr} as a function of star brightness for the targets and check stars.  Under ideal circumstances the error in the photometric zero-point dominates the total error.  Intense blending or low S/N can cause the errors to be larger.  Total errors are larger for stars fainter than 18th magnitude because of lower SNR in their pixels relative to the background.  The stars with the highest errors are those significantly influenced by blending with neighbors (see Section \ref{sec:blend}).  Most of the stars brighter than 18th magnitude with errors larger than 0.1 mag can be attributed to a handful of images of lesser quality.  They represent a relatively small fraction of the measurements, with most errors in the B, V and R bands falling between 0.04 -- 0.07 mag.  Errors in the I-band are higher (most falling between 0.05 -- 0.10 mag).  This is consistent with the higher zero point errors in the I-band attributed to lower quality I-band photometry for comparison stars.

\begin{figure}
\figurenum{3}
\label{checkhist}
\includegraphics[scale=0.33]{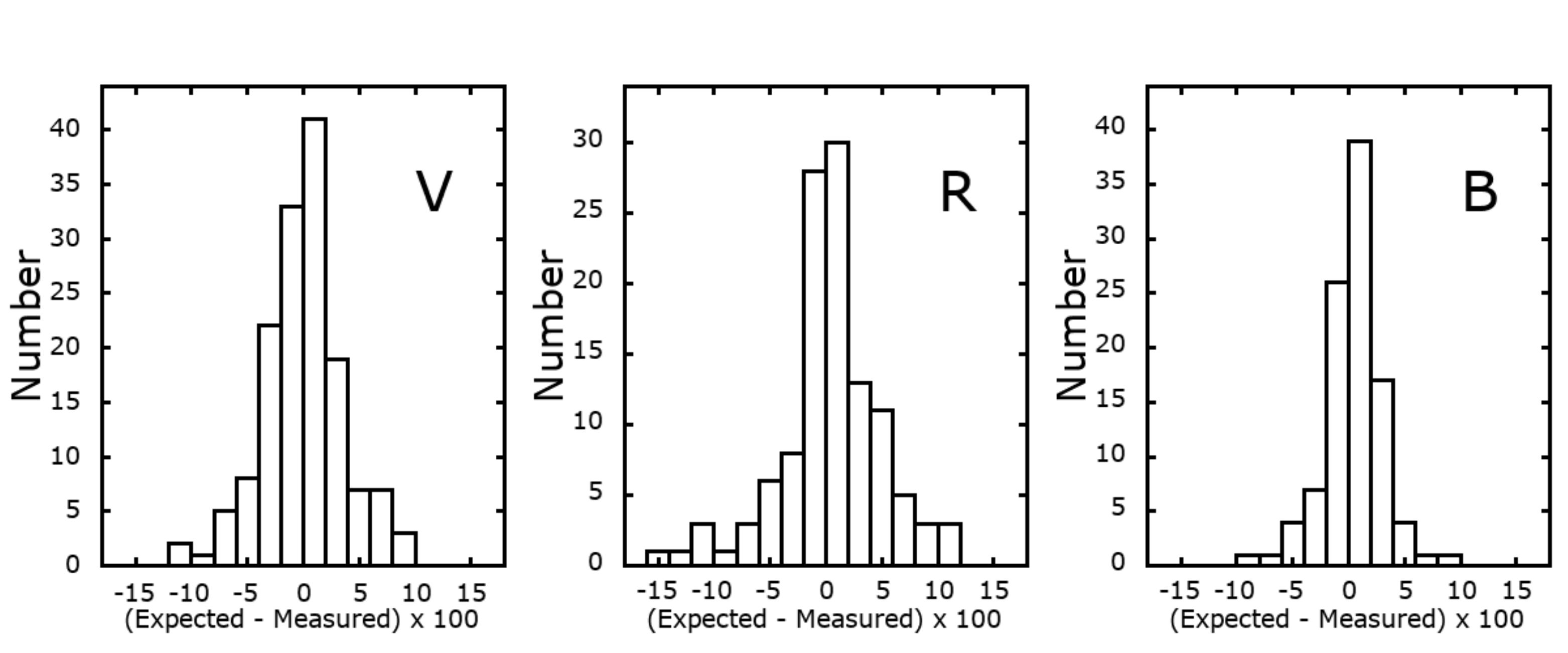}
\caption{Histograms of the magnitude difference between the measured and expected value for check stars separated by filter.}
\end{figure}

Check stars (stars of constant brightness with comparable flux relative to our targets) were identified to monitor for systematic errors in the photometric solution on individual images.  The "expected value" for each check star is the average brightness observed across all epochs.  The analysis only used check stars which were imaged in at least four epochs.  There are not yet enough observations in I-band (started in 2015) to reliably use check stars on those images.  The check stars, which tend to be fainter than the comparisons and comparable in brightness to our targets, showed no significant offsets on any of our B, V, or R images.  A histogram of the differences between the observed and expected brightness of the check stars (Figure \ref{checkhist}) confirms the random errors in our measured magnitudes are on average 0.06 magnitudes or less.  

\subsection{Blending of Targets} \label{sec:blend}

Hypergiants and classical LBVs are found in and near clusters and star forming regions \citep{2016ApJ...825...64H}.  In many cases blending can be resolved by DAOphot.  Unresolved blending impacts 105 out of 199 targets in our survey.  The concerns with unresolved blending separate into two categories:  1) unresolved blends that affect measured brightness of a target at a level equal to or less than the formal random error and 2) unresovled blends that could have an influence significantly larger than the random errors.  The LGGS catalog \citep{LGGS2} served as a valuable guide to for identifying and classifying unresolved blends.

Blends that are resolved include stars with neighbors with overlapping PSFs.  DAOphot simultaneously fits the overlapping PSFs and removes the neighbor's influence from the target.  The subtraction of the overlapping PSF is accounted for in the errors (which are modestly higher than if the star had no overlapping neighbors).  

The first case of unresolved blending includes stars which have unresolved neighbors more than 3 magnitudes fainter than the target. An unresolved blend with a star that is fainter by 3 magnitudes or more contributes no more than -0.07 mag (comparable to or less than our random errors) to a measurement.  The targets are among the most luminous stars so it is common to have faint unresolved neighbors contribute flux on the order of or less than the quoted errors.

In the second case, unresolved blends with neighbors of comparable brightness, can contribute upto -0.5 magnitudes or more to the measured flux.  It is possible for a target to have unresolved neighbors which are much fainter in one band but of comparable brightness in other bands.  For example, Var A1 in M31 is an unresolved blend with a LGGS star which is much fainter in V but comparable brightness in B, R, and I.  As a result, Var A1 is flagged as an unresolved blend of the first type in V and as the second blend type in B, R and I.    

Sometimes poor atmospheric seeing or other image quality issues can render a target unresolved from a neighbor which would be resolved under typical conditions.  If the unresolved neighbor is more than 3 magnitudes fainter, that measurement is flagged as the first type of unresolved blending.  If the temporarily unresolved neighbor is brighter, then the measurement is flagged with an "X" as unreliable (i.e. the first epoch of J004417.10+411928.0).

Targets in Table \ref{tab:goodtarg} are flagged according to their status.  A lowercase 'b' notes a minor unresolved blend (type 1) that contributes on the order of the size of the errors or less.  An uppercase 'B' denotes a more serious unresolved blend (type 2).  And uppercase 'X' denotes a measurement rendered unreliable as the result of a blend or bad PSF fit.  Notes about what is near the stars in the LGGS catalog and some rough estimates of their influence are given in Table \ref{tab:targnotes}.  Out of 72 targets in M31, 19 (26\%) are affected by blend type 1 and 8 (11\%) are affected by blend type 2.  Out of 127 targets in M33, 40 (31\%) are affected by blend type 1 and 38 (30\%) are affected by blend type 2. 

It is possible for unresolved neighbors to cause low levels of variability that should not be attributed to the target itself.  We advise discretion using this data with careful consideration given to the precision required for an analysis or application and how potential blends affect each target.  

\section{Variability of Targets} \label{sec:discussion}

The survey is designed to measure changes in target brightness over several years.  Considering the precision of the measurements, variability with an amplitude greater than 0.10-0.20 magnitudes will be detected by our survey.  We assessed the variability of each target by analyzing the correlation of brightness fluctuations in separate filters.  The limits employed in our assessment are intended to minimize false detection and provide a objective result with some bias in favor of brighter targets with smaller errors in individual measurements.

The linear correlation coefficient ($R^2$) was calculated comparing the BV, VR and BR band pairs for each target.  We considered only those targets with at least four epochs of pairs to correlate.  Because of that restriction, I-band images were not used in this analysis.  One hundred twenty (120) of the targets have at least four epochs of paired observations in all three of the bands.  An additional thirty seven (37) stars had enough epochs in two of the three band pairs. The calculated $R^2$ factors and magnitude vs magnitude plots are available for each target as part of the hyper-linked version of the catalog at \href{http://go.uis.edu/m31m33photcat}{http://go.uis.edu/m31m33photcat}.

Targets exhibiting correlation between bands that could be variability were classified into two bins:  "very likely" or "likely."  Thirteen (13) targets considered "very likely" to be variable have an $R^2$ value greater than 0.70 in all three band pairs.  An additional twenty (20) targets that exhibited an $R^2$ value greater than 0.70 in at least two band pairs were considered "likely" to be variable.  Most of the "likely" variables had some correlation ($0.70<{R^2}<0.50$) in the other band pairs but either had too few epochs with enough paired measurements and/or large errors in individual brightness measures which prevented them from being considered "very likely."   Human inspection confirmed variability in all the light curves of the targets classified "likely" and "very likely" (Table \ref{tab:var}). 

Table \ref{tab:varsum} summarizes the detected variables by spectral classification \citep{paper4}.  Known classical LBVs have the largest incidence of detected variability via band correlation (78\%).  Intermediate spectral types (YSG and warm hypergiants \citep{paper3}) have an higher incidence of detected variability (20\%-23\%) compared with earlier spectral types (OB supergiants and Of/WN stars = 0\%-11\% ).  It is unclear if this is significant since the temporal sampling rate, magnitude errors and bias against detection of variability in fainter targets must be considered to properly interpret this result.

Figure \ref{allmagerr} shows an increase in the measurement error for targets fainter than 18.0 that becomes much steeper for targets fainter than 19.0.  The 54 survey targets with an average V fainter than 18.0 include only three variables (9\% of the 33 identified).  The addition of more epochs to the survey should increase the number variables found in the sample.  With only four years of data, the larger measurement errors significantly influence our capability to detect faint variables. This bias has a roughly equal effect on the samples of OB supergiants, YSG, Of/late-WN, and warm hypergiants (see the median average V brightness and number of targets fainter than 19.0 in V in Table \ref{tab:varsum}).  However, it influences the B[e] supergiant stars (B[e]sg) more notably.

B[e]sg stars account for half of the targets with an average V brightness fainter than 19.0.  B[e]sg as a group tend to be less luminous than classical LBVs and candidate LBVs even considering their infrared flux \citep{paper4}.  No other class of star in our survey is as affected by the bias against finding faint variables.  For comparison with the other classes a summary of B[e]sg stars brighter than 19.0, having a median V brightness more comparable to the other samples, is included in Table \ref{tab:varsum}.  20\% of the bright B[e]sg are identified as possible variables.  

The candidate LBVs are stars which share spectral and physical characteristics with the quiescent state of known classical LBVs, but have not yet exhibited a period of significant brightening and spectral change associated with a classic S Doradus eruption.   Var C in M33 \citep{VarC} and M31-004526.62 \citep{newlbv} have erupted during the period of the survey.  There is also evidence that AF And began an outburst between 2015 -- 2017 \citep{afandatel,afandspec}.  We are confident that this survey is capable of detecting S Dor variability in any of the candidates which would qualify them as a definite classical LBV.  Many LBVs also exhibit Alpha Cygni type variability outside of their eruptive state (see \citet{alphacygni} and references therein).  Most of the classical LBVs not in outburst and at least three of the candidate LBVs clearly exhibit this behavior.  

\section{Conclusions And Future Work}

The first four years of our long-term survey of the most luminous stars in M31 and M33 has produced multiepoch photometry for 199 targets in four broad-band filters with 0.05 - 0.10 mag precision.  Using the correlation between bands we detected 33 possible variables, including at least two classical LBVs in eruption.  The addition of more epochs to will increase our sensitivity to detect long-term variability.

With four years completed and the fifth starting soon, our survey is the longest duration survey of M31 and M33 to date.  We will continue to image fields containing our targets at least once a year for the foreseeable future.  As the number of epochs in the survey grow we will confirm the variability of more luminous stars and build a baseline of their photometric behavior.  The data are available online at \href{http://go.uis.edu/m31m33photcat}{http://go.uis.edu/m31m33photcat}.

The images from our survey could be mined for other targets of interest.  The average limiting magnitude corresponds with an absolute magnitude of about -4 in M31 and M33.  For example, our survey could supplement the \citet{2008ApJ...684.1336K} survey of 27 nearby galaxies (excluding M31 and M33) to detect the sudden dimming of RSGs associated with failed supernovae.  Using our images the search could be extended to most of M33 and a large fraction of M31 in the visual bands as faint as $m_V\sim19$ (covering RSGs with greater than 15 $M_{\odot}$).

\section{Acknowledgments}
This work was initiated under and supported by NSF grant AST- 1108890 with additional support from the University of Illinois Springfield Henry R. Barber Astronomy Endowment funded by the people of Central Illinois.  We thank and acknowledge the UIS Observatory volunteers who helped with the imaging for the survey including:  Jim O'Brien, Jennifer Hubble-Thomas, Kevin Cranford, Greg Finn, John Lord, and Mary Sheila Tracy.  This research was also made possible through the use of the AAVSO Photometric All-Sky Survey (APASS), funded by the Robert Martin Ayers Sciences Fund.

\software{astronometry.net \citep{2010AJ....139.1782L}, DAOPHOT \citep{1987PASP...99..191S}, gnuplot (\url{http://gnuplot.info/}), IRAF \citep{1986SPIE..627..733T}}

\section{Data Tables}

\startlongtable
\begin{deluxetable*}{ccCcrCC}
\tablecaption{Image Log \label{tab:obs}}
\tabletypesize{\scriptsize}
\tablecolumns{7}
\tablenum{1}
\tablewidth{0pt}
\tablehead{
\colhead{UT\tablenotemark{a}} &
\colhead{MJD\tablenotemark{a}} &
\colhead{PSF FWHM} & \colhead{Filter} & \colhead{Field Center}&
\colhead{Rotation\tablenotemark{b}} & \colhead{Faint Limit\tablenotemark{c}} \\
\colhead{(YYYY-mm-dd)} & \colhead{(d)} &
\colhead{(arcsec)} & \colhead{} && \colhead{(Degrees)}& \colhead{(mag)}
}
\startdata
\multicolumn{7}{c}{\emph{M31}}\\
2012-08-02&56142.347& 3.0& V & 00:43:53.53 +41:55:03.2 &176.5&19.1\\
2012-08-02&56142.306& 3.6& V & 00:43:29.53 +41:41:14.9 &176.7&19.2\\
2012-08-02&56142.316& 3.2& V & 00:44:52.93 +41:35:31.9 &176.6&19.2\\
2012-08-02&56142.327& 4.0& V & 00:44:05.15 +41:16:57.1 &176.6&18.2\\
2012-08-02&56142.336& 3.7& V & 00:42:58.05 +41:04:22.1 &176.6&18.2\\
2012-12-05&56267.174& 3.1& R & 00:43:49.75 +41:54:18.9 &175.8&19.4\\
2012-12-05&56267.156& 3.1& R & 00:43:29.62 +41:41:07.0 &175.9&19.3\\
2012-12-05&56267.147& 3.1& R & 00:44:56.76 +41:34:55.4 &175.9&19.4\\
2012-12-05&56267.138& 3.2& R & 00:44:05.50 +41:16:49.0 &176.0&19.3\\
2012-12-05&56267.165& 3.4& R & 00:42:57.37 +41:03:48.6 &176.0&19.4\\
2013-07-17&56490.355& 3.1& B & 00:43:28.87 +41:41:23.1 &177.6&18.9\\
2013-07-17&56490.347& 2.6& V & 00:43:28.87 +41:41:23.3 &177.5&19.2\\
2013-07-17&56490.373& 3.4& B & 00:44:51.41 +41:35:25.9 &177.5&18.9\\
2013-07-17&56490.366& 3.0& V & 00:44:51.39 +41:35:25.8 &177.5&19.1\\
2013-07-17&56490.383& 3.2& V & 00:43:57.83 +41:17:39.6 &177.5&19.2\\
\enddata
\tablenotetext{a}{At exposure start.}
\tablenotetext{b}{Orientation of ascending row axis measured east of north in degrees. The rotation of the images changed year to year as a result of wear on the adapter holding the camera to the telescope.}
\tablenotetext{c}{A rough completeness limit estimated from peak of the magnitude frequency distribution for the image.  It may be possible to measure some stars as much as 0.5--1.0 magnitude fainter on the image.}
\tablenotetext{d}{These three images in May 2017 were taken under twilight conditions with a bright sky.}
\tablecomments{All exposures were 600 seconds in duration.}
\tablecomments{Table \ref{tab:obs} is published in its entirety in the machine-readable format.  A portion is shown here for guidance regarding its form and content.}
\end{deluxetable*}

\startlongtable
\begin{deluxetable*}{RRCCCC}
\tablecaption{Photometric Comparison Stars\label{tab:comps}}
\tabletypesize{\scriptsize}
\tablecolumns{6}
\tablenum{3}
\tablewidth{0pt}
\tablehead{
\colhead{RA\tablenotemark{a}}& \colhead{Dec\tablenotemark{a}}&
\colhead{V} & \colhead{(B-V)} & \colhead{(V-R)}& \colhead{(V-I)}
}
\startdata
\multicolumn{6}{c}{\emph{M31}}\\
9.8167&40.4346&12.106$\pm$0.024&0.366$\pm$0.035&0.225$\pm$0.043&0.439$\pm$0.060\\
9.8177&40.4869&14.679$\pm$0.027&0.670$\pm$0.040&0.389$\pm$0.042&0.755$\pm$0.059\\
9.8509&40.4435&15.955$\pm$0.030&0.706$\pm$0.055&0.394$\pm$0.058&0.763$\pm$0.083\\
9.8800&40.4643&12.690$\pm$0.022&0.307$\pm$0.032&0.189$\pm$0.042&0.370$\pm$0.060\\
9.9313&40.6618&12.895$\pm$0.022&0.733$\pm$0.033&0.430$\pm$0.043&0.833$\pm$0.061\\
9.9352&40.7567&14.302$\pm$0.029&0.657$\pm$0.043&0.402$\pm$0.060&0.779$\pm$0.085\\
9.9451&40.6834&14.714$\pm$0.029&0.624$\pm$0.043&0.384$\pm$0.050&0.745$\pm$0.071\\
9.9606&40.5858&14.483$\pm$0.032&0.683$\pm$0.049&0.428$\pm$0.033&0.829$\pm$0.046\\
9.9760&40.4870&13.010$\pm$0.019&0.664$\pm$0.032&0.397$\pm$0.040&0.770$\pm$0.057\\
\enddata
\tablenotetext{a}{J2000 Right Ascension and Declination in degrees}
\tablecomments{Table \ref{tab:comps} is published in its entirety in the machine-readable format.  A portion is shown here for guidance regarding its form and content.}
\end{deluxetable*}

\startlongtable
\begin{deluxetable*}{lclclclcl}
\tablecaption{Target Photometry\label{tab:goodtarg}}
\tablecolumns{9}
\tablenum{4}
\tablewidth{0pt}
\tablehead{
\colhead{MJD}& \colhead{V}&&
\colhead{B} && \colhead{R} && \colhead{I}&
}
\startdata
\multicolumn{9}{c}{\emph{AF And (J004333.09+411210.4)}}\\
  56142.33&17.405$\pm$0.052 &Tb&--&-&--&-&--&-\\
  56267.15&--&-&--&-&17.010$\pm$0.072 &Tb&--&-\\
  56490.38&16.963$\pm$0.031 &Tb&--&-&--&-&--&-\\
  56536.26&16.923$\pm$0.071 &b&--&-&16.713$\pm$0.074 &b&--&-\\
  56537.18&16.991$\pm$0.055 &Ab&17.071$\pm$0.049 &Ab&16.842$\pm$0.076 &b&--&-\\
  56915.16&17.305$\pm$0.035 &b&17.367$\pm$0.034 &b&17.111$\pm$0.047 &b&--&-\\
  57227.34&17.147$\pm$0.036 &b&17.204$\pm$0.033 &b&16.928$\pm$0.046 &b&--&-\\
  57310.10&17.005$\pm$0.034 &b&--&-&--&-&16.809$\pm$0.068 &b\\
  57633.35&16.709$\pm$0.058 &Ab&16.832$\pm$0.059 &Ab&16.440$\pm$0.082 &Ab&--&-\\
  57635.35&16.725$\pm$0.050 &b&--&-&--&-&16.336$\pm$0.072 &b\\
  57638.25&16.735$\pm$0.058 &b&--&-&--&-&16.371$\pm$0.136 &b\\
  57888.40&16.115$\pm$0.056 &b&16.217$\pm$0.093 &b&15.744$\pm$0.067 &b&--&-\\
\enddata
\tablecomments{Table \ref{tab:goodtarg} is published in its entirety in the machine-readable format.  A portion is shown here for guidance regarding its form and content.}
\tablecomments{Flags after each magnitude note the following:  T = magnitude not transformed because no valid color at that epoch,  A = two measurements averaged into the reported value (see text), b = blending may impact the reported magnitude by less than the reported error, B = blending may impact the reported magnitude by more than the reported error but there is a good chance that the measurement may still be reliable, X = poor image quality renders this measurement unreliable.}
\end{deluxetable*}

\startlongtable
\begin{deluxetable*}{llp{3in}}
\tablecaption{Target Notes\label{tab:targnotes}}
\tabletypesize{\scriptsize}
\tablecolumns{3}
\tablenum{5}
\tablewidth{7in}
\tablehead{
\colhead{LGGS ID}& \colhead{Other Identifier}&
\colhead{Notes}
}
\startdata
J004333.09+411210.4&AF And& Four stars $>$5 magnitudes fainter within 5 arcseconds.\\
J004334.50+410951.7&M31-004334.50& One LGGS star one mag fainter 4 arc-seconds \
to NE.  Sixteen more LGGS stars $>2$ magnitudes fainter within 5 arc-seconds.  \
On the edge of an unresolved blob but nothing bright and close enough to overwh\
elm PSF fitting.\\
J004337.16+412151.0&M31-004337.16& Seven LGGS stars $>$4 magniutdes fainter wit\
h 5 arcseconds.\\
J004339.27+411019.3&M31-004339.28&No blend concerns.\\
J004341.84+411112.0&M31-004341.84& One LGGS star 2 magnitudes fainter 4 arcseco\
nds to S does not interfere sigificantly with PSF fit.  Seven other LGGS stars \
$>$3.5 magnitudes fainter within 5 arcseconds.\\
J004350.50+414611.4&M31-004350.50& No blend concerns.\\
\enddata
\tablecomments{Table \ref{tab:targnotes} is published in its entirety in the machine-readable format.  A portion is shown here for guidance regarding its form and content.}
\end{deluxetable*}

\clearpage

\startlongtable
\begin{deluxetable*}{llCcCCCC}
\tabletypesize{\scriptsize}
\tablecaption{Possible Variables \label{tab:var}}
\tablecolumns{8}
\tablenum{6}
\tablewidth{7in}
\tablehead{
&&\colhead{Average}&&\colhead{Range}&\multicolumn{3}{c}{$R^2$ Correlation\tablenotemark{a}}\\
\colhead{LGGS ID}&\colhead{Name}&\colhead{V Mag}&\colhead{Var?}&\colhead{V Mag}&\colhead{BV}&\colhead{VR}&\colhead{BR}
}
\startdata
\multicolumn{8}{c}{OB Supergiants}\\
J004434.64+412503.5&M31-004434.65&16.68&likely&0.11&0.89 (4)&0.87 (4)&-\\
J013309.14+304954.5&M33C-23048&17.98&likely&0.10&0.94 (5)&0.63 (5)&0.81 (4)\\
J013312.81+303012.6&M33C-4119&17.47&very likely&0.55&0.86 (6)&0.96 (7)&0.82 (6)\\
J013339.52+304540.5&M33C-19725&17.27&very likely&0.51&0.95 (6)&0.96 (6)&0.96 (5)\\
J013437.25+303817.7&V-139873&17.48&likely&0.13&0.74 (7)&0.88 (8)&0.51 (6)\\
\multicolumn{8}{c}{Warm Hypergiants}\\
J013352.42+303909.6&V-093351&16.44&likely&0.51&0.91 (9)&0.89 (9)&0.64 (7)\\
J013415.42+302816.4&V-125093&17.42&likely&0.16&0.42 (5)&0.88 (4)&0.93 (4)\\
\multicolumn{8}{c}{Yellow Supergiants}\\
J004424.21+412116.0&M31-004424.21&16.71&likely&0.25&0.94 (4)&0.94 (4)&-\\
J004507.65+413740.8&M31-004507.65&16.16&likely&0.14&0.90 (4)&0.95 (4)&-\\
J013303.4+303051.5&M33-013303.40&17.36&likely&0.44&0.78 (7)&0.94 (8)&-\\
J013303.6+302903.4&M33-013303.60&17.19&likely&0.18&0.81 (6)&0.91 (7)&0.53 (6)\\
J013355.78+304831.3&M33C-22178&19.35&very likely&0.27&0.83 (5)&0.88 (4)&0.76 (4)\\
J013358.96+304139.5A&V-104139&16.86&likely&0.26&0.73 (10)&0.77 (10)&0.35 (8)\\
J013359.37+302310.9&V-104958&16.94&very likely&0.26&0.93 (5)&0.78 (6)&0.71 (5)\\
J013420.65+303942.6&M33C-14120&18.19&very likely&0.36&0.77 (7)&0.78 (8)&0.78 (6)\\
\multicolumn{8}{c}{Classical LBVs}\\
J004333.09+411210.4&AF And&16.99&very likely&0.70&1.00 (4)&0.98 (5)&0.97 (4)\\
J004419.43+412247.0&Var 15&17.91&likely&0.70&1.00 (4)&1.00 (4)&-\\
J004450.54+413037.7&Var A1&16.94&likely&0.35&0.98 (4)&0.99 (4)&-\\
J004526.62+415006.3&M31-004526.62&16.53&likely&0.44&0.97 (5)&0.97 (5)&-\\
J013335.14+303600.4&Var C&15.74&very likely&0.82&0.93 (8)&1.00 (8)&0.95 (8)\\
J013349.23+303809.1&Var B&17.32&very likely&0.47&0.83 (7)&0.93 (7)&0.90 (6)\\
J013410.93+303437.6&Var 83&16.06&very likely&0.71&0.95 (10)&0.99 (11)&0.95 (9)\\
\multicolumn{8}{c}{B[e]sg}\\
J004417.10+411928.0&M31-004417.10&17.12&likely&0.29&0.90 (4)&0.75 (4)&-\\
J013242.26+302114.1&M33-013242.26&18.12&very likely&0.36&0.85 (4)&0.77 (5)&0.96 (4)\\
\multicolumn{8}{c}{Canidate LBVs}\\
J013235.25+303017.6&M33C-4174&17.86&likely&0.08&0.85 (4)&0.80 (5)&0.61 (4)\\
J004425.18+413452.2&M31-004425.18&17.38&likely&0.38&0.93 (4)&0.92 (4)&-\\
J013432.74+304709.6&M33C-21192&15.96&likely&0.32&0.93 (4)&0.73 (4)&-\\
\multicolumn{8}{c}{Unknown Classification}\\
J003907.59+402628.4&M31-003907.59&16.77&very likely&0.19&0.94 (5)&0.90 (5)&0.74 (5)\\
J004021.21+403117.1&M31-004021.21&16.99&likely&0.13&0.69 (7)&0.93 (7)&0.82 (8)\\
J004034.82+401825.5&M31-004034.82&16.34&very likely&0.08&0.85 (5)&0.96 (5)&0.86 (6)\\
J013237.72+304005.6&UIT003&17.81&likely&0.58&0.93 (4)&0.99 (4)&-\\
J013351.00+303818.8&M33C-12559&16.73&very likely&0.40&0.86 (8)&0.90 (8)&0.81 (7)\\
\enddata
\tablenotetext{a}{Number of epochs in calculated correlation given in parentheses.  Magnitude pairs with fewer than 4 epochs are not considered.}
\tablenotetext{b}{Spectral classification from \citet{paper4}.}
\end{deluxetable*}

\clearpage

\startlongtable
\begin{deluxetable*}{lrcCCccc}
\tabletypesize{\scriptsize}
\tablecaption{Variability Detected In Each Class\tablenotemark{a}\label{tab:varsum}}
\tablecolumns{7}
\tablenum{7}
\tablewidth{7in}
\tablehead{
&&&\colhead{Median}&\colhead{Percent}&\multicolumn{3}{c}{Detection Bin} \\
\colhead{Class\tablenotemark{a}}&\colhead{N}&\colhead{N(V$>$19.)}&\colhead{V mag}&\colhead{Variable}&\colhead{Very Likely}&\colhead{Likely}
}
\startdata
Of/late-WN&15&0&17.95&0\%&0&0\\
OB Supergiants&47&1&17.67&11\%&2&3\\
Yellow Supergiants&35&2&17.36&23\%&3&5\\
Warm Hypergiants&10&1&18.12&20\%&0&2\\
Classical LBVs&9&0&16.99&78\%&4&3\\
\hline
Candidate LBVs\tablenotemark{b}&23&1&17.86&23\%&0&3\\
B[e] Supergiants (All)&18&8&18.85&11\%&1&1\\
B[e] Supergiants (V$<$19)&10&0&18.17&20\%&1&1\\
\hline
Unknown&19&3&17.75&26\%&3&2\\
Peculiar&3&0&17.11&0\%&0&0\\
\enddata
\tablenotetext{a}{Spectral classification from \citet{paper4}.}
\tablenotetext{b}{Stars having spectral characteristics of classical LBVs that have never been observed in eruption. (Table 7 in \citet{paper4}).}
\end{deluxetable*}



\end{document}